\documentstyle[12pt]{article}
\newcommand{\ald}{\dot \alpha}
\newcommand{\bed}{\dot \beta}
\newcommand{\gad}{\dot \gamma}

\newcommand{\dald}{{\bar{\partial}}_{\dot \alpha}}
\newcommand{\nald}{{\bar{\nabla}}_{\dot \alpha}}
\newcommand{\ned}{{\bar{\nabla}}_{\dot \beta}}

\newcommand{\ej}{\cal E}
\newcommand{\hej}{\hat {\ej}}
\begin{document}
\textwidth 160mm
\textheight 240mm
\topmargin -20mm
\oddsidemargin 0pt
\evensidemargin 0pt
\newcommand{\beq}{\begin{equation}}
\newcommand{\eeq}{\end{equation}}
\begin{titlepage}
\begin{center}

\huge{\bf On amplitudes in self-dual sector of Yang-Mills theory}

\vspace{1.5cm}

\large{\bf A.A.Rosly and  K.G.Selivanov }

\vspace{1.0cm}

{ITEP,B.Cheremushkinskaya 25, Moscow,117259, Russia}

\vspace{1.9cm}

{ITEP-TH-50/96}

\vspace{1.0cm}

\end{center}
              

\begin{abstract}
Self-dual perturbiner in the Yang-Mills theory is constructed by the twistor methods both
in topologically trivial and topologically nontrivial cases.  Maximally helicity violating 
amplitudes and their instanton induced analogies are briefly discussed.
\end{abstract}
\end{titlepage}

\newpage
\setcounter{equation}{0}

1.The tree and one-loop amplitudes in the self-dual (SD) sector of the
Yang-Mills (YM) theory, also known as like-helicity multi-gluon amplitudes,
have been intensively studied in the literature (see, e.g.,  reviews 
\cite{mapa}, \cite{dixon} and  refs. \cite{AM1}-\cite{AM10} where the
amplitudes were discussed from some different points of view). From our point of view,
such amplitudes provide one of the most interesting illustrations of the 
concept of {\it perturbiner} introduced in ref.\cite{RS}. The 
perturbiner is a generating function
for (a class of) tree amplitudes in the theory. As explained in ref.\cite{RS},
the perturbiner can be given intrinsic definition which is formally independent 
of considering Feynman diagrams. It is a solution of the field equations
which uniquely corresponds to a given plane waves solution of the free (zero coupling
constant) field equations. The set of plane waves included in the solution of the 
free field equations is essentially the set of asymptotic states in the amplitudes
which the perturbiner is the generating function for. Considering the asymptotic states of only
negative helicity leads to considering only SD solutions of the YM equations.

In this letter we construct the SD perturbiner
for the YM theory with arbitrary gauge group. 
We find a first order anti-SD deformation of the SD perturbiner
which allows us to describe amplitudes with any number of gluons from the SD
sector and with two gluons from the anti-SD sector. We find fermionic deformations
of the SD perturbiner. The possibility to have 
an arbitrary instanton background is also taken into account. 

In  the topologically trivial
case the SD perturbiner reproduces the prominent expressions from
refs.\cite{PT}, \cite{BG} for the like-helicity amplitudes (or, rather, for the 
like-helicity ``currents'', i.e. objects with a number of on-shell same
helicity gluons and one arbitrary of-shell gluon).
This type of SD solutions has been discussed
in refs.\cite{Ba}, \cite{Se}, \cite{KO}. In ref.\cite{Ba} and independently
in ref.\cite{Se}, the tree like-helicity amplitudes were related to solutions
of the SD equations. In \cite{Ba} it was basically shown that  the SD
equations reproduce the recursion relations obtained originally in ref.\cite{BG}
from the Feynman diagrams; the  corresponding solution of SD equations
was obtained with use of the known solution of refs.\cite{PT}, \cite{BG}
of the recursion relations. In ref.\cite{Se} an example of SD perturbiner was
obtained in the SU(2) case by a 'tHooft anzatz upon further restriction on the
asymptotic states included. The consideration of ref.\cite{KO} is based
on solving  recursion relations analogous to refs. \cite{BG}.

In the topologically
nontrivial case the SD perturbiner generates instanton mediated negative helicity
amplitudes, closed expressions for which, as far as we know, have been absent in the literature.
A particular example of the topologically nontrivial SD perturbiner for the
SU(2) case was described in ref.\cite{RS}. A topologically nontrivial perturbiner 
in 2D K\"ahler sigma model has been considered in ref.\cite{RS}.
\\
\\
2.  We adopt the spinor notations, i.e., $A_{\alpha \ald}, \alpha=1,2, \ald={\dot 1}, {\dot 2}$ 
stands for the YM potential.
The perturbiner is a complex solution of the field equations. 
In the spinor notations the curvature form, $F=dA+A^{2}$,
has four indices, $F_{\alpha \ald \beta \bed}$, and, being antisymmetric with respect to 
permutation $(\alpha \ald)\leftrightarrow(\beta \bed)$, decomposes as
\beq
\label{decom}
F_{\alpha \ald \beta \bed}=\varepsilon_{\alpha \beta}
F_{\ald \bed}+\varepsilon_{\ald \bed} F_{\alpha \beta}
\eeq
where $\varepsilon_{\alpha \beta}, \varepsilon_{\ald \bed}$ are the usual antisymmetric 
tensors.
 The first term in the r.h.s. of Eq.~(\ref{decom})
can be identified as an SD component of $F$, the second one - as an anti-SD component
of $F$. Correspondingly, the SD equation can be written as
\beq
\label{SD}
F_{\alpha  \beta}=0
\eeq
and the anti-SD equation - as
\beq
\label{anti-SD}
F_{\ald \bed}=0
\eeq
The free SD (anti-SD) equation is the same as Eq.~(\ref{SD}) (Eq.~(\ref{anti-SD}))
with $F=dA+A^{2}$ substituted by $F^{(0)}=dA^{(0)}$. A free solution consisting
of $N$ plane SD-waves is as follows
\beq
\label{free}
A^{(0) N}_{\alpha \ald}=\sum_{j}^{N}A_{\alpha \ald}(j)
\eeq
where the sum runs over gluons, $N$ is the number of gluons,\newline
\mbox{$A_{\alpha \ald}(j)=i a_{j} t_{j} \epsilon^{+j}_{\alpha \ald} 
e^{i k^{j}_{\alpha \ald} x^{\alpha \ald}}$},
$x^{\alpha \ald}$ represents the space-time coordinate, $k^{j}_{\alpha \ald}$ is a 
light-like four momentum of the $j$-th
gluon, $\epsilon^{+j}_{\alpha \ald}$ is a four-vector defining a polarization of the $j$-th gluon,
$t_{j}$ is a matrix defining colour orientation of the $j$-th gluon, 
$a_{j}$ is the symbol of annihilation/creation operator (depending on the sign of the time
component of $k^{j}_{\alpha \ald}$) of the $j$-th gluon. As in ref.\cite{RS}, we assume 
nilpotency of $a_{j}$, that is $a_{j}^{2}=0$. Since $N$ is arbitrary, the nilpotency
can be assumed without any loss of generality. Below we use the following short-cut
notations
\begin{eqnarray}
\label{shcu}
{\ej}^{j}=a_{j}e^{i k^{j}_{\alpha \ald} x^{\alpha \ald}}\nonumber\\
{\hej}^{j}=t_{j}{\ej}^{j}
\end{eqnarray}
$k^{j}_{\alpha \ald}$, as a light-like four-vector, decomposes into a product of two
spinors
\beq
k^{j}_{\alpha \ald}=\ae^{j}_{\alpha} \lambda^{j}_{\ald}
\eeq
Since we are considering SD-waves, $\epsilon^{+j}_{\alpha \ald}$ is also a light-like 
four-vector of the following form 
\beq
\epsilon^{+j}_{\alpha \ald}=\frac{q^{j}_{\alpha} \lambda^{j}_{\ald}}{( \ae^{j},q^{j})}
\eeq
where normalization is defined with use of a convolution 
$(\ae^{j},q^{j} )=\varepsilon^{\gamma \delta}\ae^{j}_{\gamma}q^{j}_{\delta}=
{\ae^{j}}^{\delta}{q^{j}}_{\delta}$.

The free anti-SD equation would give rise to a polarization
$\epsilon^{-}_{\alpha \ald}$,
\beq
\epsilon^{-}_{\alpha \ald}=\frac{\ae_{\alpha} \bar{q}_{\ald}}
{( \lambda,\bar{q})}
\eeq
The auxiliary spinors $q_{\alpha}$ and $\bar{q}_{\ald}$ form together a four-vector
$q_{\alpha \ald}=q_{\alpha}\bar{q}_{\ald}$ usually called a reference momentum.
The normalization was chosen so that
\beq
\epsilon^{+} \cdot \epsilon^{-}=\varepsilon^{\alpha \beta}\varepsilon^{\ald \bed}
\epsilon^{+}_{\alpha \ald} \epsilon^{-}_{\beta \bed}=-1
\eeq

By definition (cf. ref.\cite{RS}), the perturbiner $A^{ptb}$ is a solution of the field 
equations, of the SD equations in the present case, which is polynomial in the nilpotent
symbols ${\ej}^{j}$. In the topologically trivial case the first order term of the polynomial
$A^{ptb}$ is equal to $A^{(0)}$ from Eq.~(\ref{free}), while the zeroth order term is absent.
In the topologically nontrivial case the zeroth order term of $A^{ptb}$ is equal to an instanton
field and the first order term far from the center of the instanton is the gauge transformed  
$A^{(0)}$ from Eq.~(\ref{free}) with the same gauge transformation which defines
the asymptotic of the instanton.

Below we shall assume that the pertrubiner $A^{ptb}$ obeys the Bose symmetry property 
which means that it is invariant under permutations of the type 
$({\hej}^{j_{1}}, \ae^{j_{1}}, \lambda^{j_{1}}, 
q^{j_{1}})\leftrightarrow({\hej}^{j_{2}}, \ae^{j_{2}}, \lambda^{j_{2}}, 
q^{j_{2}})$
and  the restriction property
\beq
\label{restr}
{A^{ptb}_{N}|}_{({\ej}^{N}=0)}=A^{ptb}_{N-1}
\eeq
where the subscript $N$ indicates the  number of  gluons included.
Eq.~(\ref{restr}), in particular, means that $A^{ptb}_{N-1}$ is independent
of quantum numbers of the $N$-th gluon.

One can see that such a solution  is unique up to gauge transformations (cf. ref.\cite{RS}).

Before describing the solution, we introduce some more notation. 
With use of the auxiliary twistor variables, $p^{\alpha}, \alpha=1,2$, which can be 
viewed on as a pair of complex numbers, we form objects
\begin{eqnarray}
\label{1}
A_{\ald}=p^{\alpha}A_{\alpha \ald},\nonumber\\
\dald=p^{\alpha} \partial_{\alpha \ald}, \;
{\rm where} \; \partial_{\alpha \ald}=\frac{\partial}{\partial x^{\alpha \ald}},\nonumber\\
\nald=\dald+A_{\ald}
\end{eqnarray}
The SD equation, Eq.~(\ref{SD}), then takes the form of a zero-curvature condition 
(ref.\cite{Ward})
\beq
\label{SD1}
[\nald, \ned]=0, \; {\rm at \: any} \; p^{\alpha}, \alpha=1,2
\eeq
which is solved as
\beq
\label{gauge}
A_{\ald}=g^{-1}{\dald}g
\eeq
where $g$ is a function of $x^{\alpha \ald}$ and $p^{\alpha}$ with values in the 
complexification of the
gauge group. $g$ must depend on $p^{\alpha}$ in such a way that the resulting
$A_{\ald}$ is a linear homogeneous function of $p^{\alpha}$, as in Eq.~(\ref{1}).
Probably, it is worth to stress that $A_{\ald}$ from Eq.~(\ref{gauge})
is not necessary a pure gauge since $g$ is $p^{\alpha}$-dependent.

Below we explicitly construct such a matrix $g^{ptb}$ for the SD perturbiner. Notice
that $g^{ptb}$ inherits the Bose symmetry and restriction properties.
\\
\\
3. In the topologically trivial sector the perturbiner depends on $x^{\alpha \ald}$
only via ${\hej}^{j}, j=1, \ldots ,N$, so that instead of an infinite-dimentional
function space we deal with a  
 finite-dimensional space of polynomials in $N$ nilpotent variables.
Using the nilpotency of ${\hej}^{j}$ and the restriction property Eq.~(\ref{restr})
one can see that 
\beq
\label{induc}
g^{ptb}_{N}=g^{ptb}_{N-1}(1+\chi_{N})
\eeq
where $\chi_{N}$ is of first order in ${\hej}^{N}$ and polynomial in all ${\hej}^{j}$.
Here $g^{ptb}$ is assumed to be a rational function of $p^{\alpha}$.
The linear part of $\chi_{N}$ is fixed by the condition that the linear part of $A^{ptb}$
is $A^{(0)}$ from Eq.~(\ref{free}) and the higher order terms in $\chi_{N}$ are fixed by
demanding regularity of $A^{ptb}$ as a function of
$p^{\alpha}, \alpha=1,2$. More precisely, the above conditions fix $\chi_{N}$ up to
a freedom which is equivalent to the gauge freedom. There is, however, a minimal
choice for $\chi_{N}$ which happens to correspond to the Lorentz gauge for $A^{ptb}$.
Upon that choice one gets
\beq
\label{res4}
 \chi_{N}=\frac{(p,q^{N})}{(p, \ae^{N})}h_{N-1}^{-1}
 \frac{{\hat {\cal E}}^{N}}{(\ae^{N},q^{N})}h_{N-1}
\eeq
where we have introduced $h_{N-1}={g^{ptb}_{N-1}|}_{(p=\ae^{N})}$. With 
$\chi_{N}$ from Eq.~(\ref{res4}), Eq.~(\ref{induc}) becomes a recursion relation for 
$g^{ptb}$. This recursion
relation greatly simplifies if one considers ordered \footnote{One can see that the decomposition
of any polynomial in ${\hej}^{j}$'s into the ordered monomials is well defined.}
highest degree monomials in $g^{ptb}$, say,
\beq
\label{order}
g^{ptb}_{N (N, \ldots ,1)}=C^{j}(\ae, q) \hej^{N} \ldots \hej^{1}
\eeq
for which the recursion relation readily leads to
\beq
\label{bsol}
g_{N (N, \ldots ,1)}=\frac{(p, q^{N})(\ae^{N}, q^{N-1}) \ldots 
(\ae^{2}, q^{1})}{(p, \ae^{N})(\ae^{N}, \ae^{N-1}) \ldots
(\ae^{2}, \ae^{1})}
\frac{ {\hej}^{N}}{( \ae^{N},q^{N})} \ldots \frac{{\hej}^{1}}{( \ae^{1},q^{1})}
\eeq
This is, essentially, a solution of the problem. The whole $g^{ptb}$ is restored
with use of the Bose symmetry and the restriction property,
\beq
\label{result}
g^{ptb}=\sum_{d=0} \sum_{J_{d}}\frac
{(p, q^{j_{d}})(\ae^{j_{d}}, q^{j_{d-1}}) \ldots (\ae^{j_{2}}, q^{j_{1}})}
{(p, \ae^{j_{d}})(\ae^{j_{d}}, \ae^{j_{d-1}}) \ldots 
(\ae^{j_{2}}, \ae^{j_{1}})} 
\frac{\hej^{j_{d}}}{( \ae^{j_{d}},q^{j_{d}})} \ldots 
\frac{\hej^{j_{1}}}{(\ae^{j_{1}},q^{j_{1}})}
\eeq
where the second sum runs over all ordered subsets $J_{d}=\{j_{1}, \ldots ,j_{d}\}$ of the set 
$\{1, \ldots , N-1\}$.

Substituting $g^{ptb}$ from  Eq.~(\ref{result}) into Eq.~(\ref{gauge}) determines 
the perturbiner
$A^{ptb}_{\ald}$. At first glance, the corresponding computation looks somewhat 
cumbersome, but again there is a short-cut. By construction, $A^{ptb}_{\ald}$ is a linear 
homogeneous function of $p^{\alpha}$. Finding such a function is equivalent to finding
its derivative with respect to $p^{\alpha}$ at any value of $p^{\alpha}$. We put at this 
moment all $q$'s equal each 
other (it is a gauge 
choice for the gluons) and compute the derivative of $A^{ptb}_{\ald}$ at $p=q$. 
In this case $g^{ptb}|_{(p=q)}=1$, and the perturbiner sought-for is
\beq
\label{A}
A^{ptb}_{\alpha \ald}
=i \sum_{d=1} \sum_{J_{d}} 
\frac{(\sum_{l  \in {J_{d}}} q_{\alpha}( \ae^{l},q) 
{\lambda}^{l}_{\ald})}
{(\ae^{j_{d}},q)(\ae^{j_{1}},q)}
 \frac{{\hej}^{j_{d}} \ldots 
{\hej}^{j_{1}}}{( \ae^{j_{d}}, \ae^{j_{d-1}}) \ldots 
(\ae^{j_{2}}, \ae^{j_{1}})} 
\eeq
\\
\\
4. The SD perturbiner can be used as a base point for a perturbation 
procedure of adding one-by-one gluons of the opposite helicity, or other particles,
say, fermions, interacting with gluons (cf. ref.\cite{RS}, \cite{Se}). 
The explicit expression for $g^{ptb}$, Eq.~(\ref{result}), obtained above is very useful 
in this procedure.

The SD perturbiner 
itself describes the so-called off-shell currents - objects including an arbitrary number of
on-shell SD gluons and one arbitrary off-shell gluon. When the latter gluon becomes on-shell,
one gets an amplitude with all gluons but one having the same helicity and the latter one 
having an arbitrary helicity. Such amplitudes are known to be zero 
(see, e.g., the review \cite{mapa}). That vanishing can be seen applying to the perturbiner 
Eq.~(\ref{A}) the reduction formula ref.\cite{RS}, \cite{Se}
\beq
\label{reduc}
M(k'', \{a_{j}\})=-i \int d^{4}x \, tr[(dA'') \cdot (dA^{ptb})]
\eeq
where $M$ is a generating function for the amplitudes with one marked gluon
of arbitrary helicity, $k''$ stands for the quantum numbers of the marked gluon, $A''$ is
the corresponding solution of the free equations, $d$ stands for the external 
derivative, $ \cdot $ indicates the scalar product defined by the space-time metric, $tr$ is the
trace. The amplitudes
with one marked particle are generated as coefficients in expansion of $M$ in 
powers of the symbols $a_{j}$.

To include one more gluon of the opposite helicity one needs to find a first order 
anti-SD deformation of the SD perturbiner, that is, to solve the linearized YM equation 
in the background of SD perturbiner. Its solution goes as follows.

The YM equations 
\begin{eqnarray}
\nabla \ast F=0\nonumber\\
\nabla  F=0
\end{eqnarray}
rewrite as
\begin{eqnarray}
\label{YM}
\varepsilon^{\alpha \beta} \nabla_{\alpha \ald}F_{\beta \gamma}=0\nonumber\\
\varepsilon^{\ald \bed} \nabla_{\alpha \ald}F_{\bed \gad}=0
\end{eqnarray}

At first step one solves the variation of the first of Eqs.~(\ref{YM}),
\beq
\label{variation}
\varepsilon^{\alpha \beta} \nabla_{\alpha \ald}f_{\beta \gamma}=0
\eeq
where $f_{\beta \gamma}=
\delta F_{\beta \gamma}$. $f_{\beta \gamma}$
must be of the first order in ${\hej}'$ and polynomial in ${\hej}^{j}, j=1, \ldots , N$,
where ${\hej}'$ corresponds to the added anti-SD gluon. As above, the first order term,
$f^{(0)}_{\beta \gamma}$,
in the polynomial $f_{\beta \gamma}$ must be a solution of the free anti-SD equation,

\beq
\label{ffree}
f^{(0)}_{\alpha \beta}=\ae'_{\alpha}\ae'_{\beta} \hej',
\eeq
the corresponding four-momentum $k'_{\alpha \ald}$ being 
$k'_{\alpha \ald}=\ae'_{\alpha} \lambda'_{\ald}$.

At second step one finds a potential $a_{\alpha \ald}$ such that anti-SD
component of its covariant derivative is the above $f_{\alpha \beta}$
\beq
(\nabla a)_{\alpha \beta}=f_{\alpha \beta}
\eeq
where the covariant derivative $ \nabla_{\alpha \ald}$ is in the SD perturbiner
background, $A^{ptb}_{\ald}=(g^{ptb})^{-1}{\dald}g^{ptb}$ .

The first step is readily done,
\beq
\label{pata}
f_{\alpha \beta}=(g^{ptb}|_{(p=\ae')})^{-1}f^{(0)}_{\alpha \beta} 
g^{ptb}|_{(p=\ae')}
\eeq

As concernes the second step, one luckily need not doing it. Indeed, the reduction
formula relating amplitudes with two marked gluons to the deformation $a_{\alpha \ald}$
of the perturbiner reads 
\beq
\label{reduc2}
M(k'',k', \{a_{j}\})=
-i \int d^{4}x \, tr [(dA'') \cdot (da)]
\eeq
where $M$ is a generating function for such amplitudes, $k'',k'$ stand for the
quantum numbers of the marked gluons, $k'$ entering $a$, the solution of the second
step of the problem above. Assuming the doubly primed gluon to be anti-SD  Eq.~(\ref{reduc})
rewrites as
\beq
\label{redrew}
M(k'',k', \{a_{j}\})=
-i \int d^{4}x \, tr [{\hej}'' {{\ae}''}^{\alpha}{{\ae}''}^{\beta} 
\partial_{\alpha \ald} a_{\beta \bed} \varepsilon^{\ald \bed}]
\eeq
where ${\hej}''$ corresponds to the doubly primed gluon, the corresponding four momentum
$k''_{\alpha \ald}$ decomposing as $k''_{\alpha \ald}={{\ae}''}_{\alpha} 
{{\lambda}''}_{\ald} $

Taking the reference momentum, $q$, in the perturbiner $A^{ptb}$ such that 
$({\ae}'',q)=0$ one can substitute in the r.h.s. of Eq.~(\ref{redrew})
the derivative $\partial_{\alpha \ald}$ by the covariant derivative $\nabla_{\alpha \ald}$
in the $A^{ptb}$ background. This way, with use of the solution Eq.~(\ref{pata}),
one comes to a compact expression for the
generating function of the Parke-Taylor \cite{PT} amplitudes
\beq
\label{patage}
M(k'',k', \{a_{j}\})=-i ( {\ae}'', \ae')^{2}
 \int d^{4}x \, tr [{\hej}''(g^{ptb}|_{(p={\ae}')})^{-1} {\hej}'
g^{ptb}|_{(p={\ae}')}]
\eeq
In verifying the equivalence of Eq.~(\ref{patage}) to the Parke-Taylor expressions \cite{PT}
we recommend to consider the terms of definite cyclic order in ${\hej}$'s..

To include into the game a couple of fermions is as easy as to include a couple of 
anti-SD gluons, because the fermion field equations are analogous to 
Eq.~(\ref{variation}). More precisely, depending on chirality of the fermions,
the field equations look as
\beq
\label{ferri}
\varepsilon^{\alpha \beta} \nabla_{\alpha \ald} \Psi_{\beta}=0
\eeq
or as
\beq
\label{ferle}
\varepsilon^{\ald \bed} \nabla_{\alpha \ald} \Psi_{\bed}=0
\eeq
where $\nabla_{\alpha \ald}$ is in  background of the SD perturbiner. Again the solution
$\Psi_{\beta}$ ($ \Psi_{\bed}$)
must be of the first order in ${\ej}'$ and polynomial in ${\hej}^{j}, j=1, \ldots , N$,
where ${\ej}'$ belongs to the fermion, and a linear term of the polynomial
$\Psi_{\beta}$ ($ \Psi_{\bed}$) must be a solution of the free fermion equations,
$\Psi^{(0)}_{\beta}=\ae'_{\alpha}{\hej}'$ 
($ \Psi^{(0)}_{\bed}=\lambda'_{\ald}{\hej}'$). Four-momentum of the
fermion, $k'_{\alpha \ald}$, in both cases decomposes as 
$k'_{\alpha \ald}=\ae'_{\alpha} \lambda'_{ \ald}$. The hat above 
${\ej}'$ in this case
indicates that ${\hej}'$ includes a vector from the gauge group representation.
Such solutions of Eqs.~(\ref{ferri}), (\ref{ferle}) read as
\begin{eqnarray}
\Psi_{\beta}=({g^{ptb}|_{(p=\ae')}})^{-1}\Psi^{(0)}_{\beta}\nonumber\\
\Psi_{\bed}=-i\frac{q^{\alpha} \partial_{\alpha \bed}}{(q, \ae')}
({g^{ptb}|_{(p=\ae')}})^{-1}{\hej}'
\end{eqnarray}
where $g^{ptb}$ acts on ${\hej}'$ in the corresponding representation.
Using these expressions one can easily write down the so-called off-shell
fermionic current used in ref.\cite{BG}. Notice that amplitudes
with two on-shell massless fermions and any number of SD gluons vanish.
\\
\\
5. As we mentioned above (cf. ref.(\cite{RS}), the concept of perturbiner can be generalized
to a topologically nontrivial sector. In the latter case  it provides a framework for the 
instanton mediated  multi-particle amplitudes. Intuitively, the topologically nontrivial perturbiner
is a sort of hybrid of the topologically trivial perturbiner considered above and of the
standard instanton solution. Again, it is a polynomial in the same 
nilpotent variables ${\ej}^{j}$, 
but coefficients of the polynomial are now (matrix valued) functions on the Eucleadian space
(or their analytical continuations to the Minkowski space, cf. ref.\cite{RS}). At 
${\ej}^{j}=0, j=1,2, \ldots $, the topologically nontrivial perturbiner $A^{iptb}_{\ald}$ 
is just the instanton,
\beq
A^{iptb}_{\ald}|_{({\ej}^{j}=0, j=1, \ldots )}=A^{inst}_{\ald}
\eeq
(which can be considered as a particular case of the restriction property Eq.~(\ref{restr})).
All what we need to know about the instanton, $A^{inst}_{\ald}$, that it can be represented
in the twistor-spirit form
\beq
A^{inst}_{\ald}=g_{inst}^{-1} \dald g_{inst}
\eeq
$g_{inst}$ is assumed to be a rational function of  the auxiliary variables $p^{\alpha}$,
 such that $A^{inst}_{\ald}$ is a linear homogeneous function
of $p^{\alpha}$.  Then the SD topologically nontrivial perturbiner
$A^{iptb}_{\ald}$ is represented in the form 
\beq
\label{igauge}
A^{iptb}_{\ald}=({g^{iptb}})^{-1} \dald g^{iptb}
\eeq
and the corresponding $g^{iptb}$ is found to be 
\beq
\label{iptb}
g^{iptb} ({\hej}^{1}, {\hej}^{2}, { \ldots }) =g^{inst}
g^{ptb} ({{\hej}^{1}}_{g}, {{\hej}^{2}}_{g}, { \ldots } )
\eeq
where ${\hej}^{j}_{g}, j=1,2, \ldots $ stand for twisted harmonics
\beq
{\hej}^{j}_{g}=(g_{inst}|_{(p=\ae^{j})})^{-1} \hej^{j}g_{inst}|_{(p=\ae^{j})}
\eeq
and $g^{ptb}$ is defined in Eq.~(\ref{result}).

$g^{iptb}$ from Eq.(\ref{iptb}) allows one to reproduce in the instanton sector all the
results described above in the topologically trivial case. More detailed account of these issues,
as well as calculation of one-loop corrections, will be done elsewhere \cite{RS2}.
\\
\\
The work of A.A.R. was partially supported by the grants RFFR-96-02-18046 and  
INTAS-93-0166.

\end{document}